\documentclass[aps,pra,superscriptaddress,amsmath,amssymb,preprintnumbers,showpacs
]{revtex4-2}
\usepackage{amssymb}
\usepackage{bm}
\usepackage{color}

\usepackage{amsfonts}
\usepackage{amsmath}
\usepackage{amssymb,amsthm}
\usepackage{mathtools}
\usepackage{color}

\usepackage{url}
\usepackage{graphicx}
\usepackage{booktabs}
\usepackage{multirow}
\usepackage{comment}

\numberwithin{equation}{section}             

\newtheorem{theorem}{Theorem}

\newtheorem{corollary}{Corollary}
\newtheorem{remark}{Remark}
\newtheorem{proposition}{Proposition}

\begin{document}


\title{Characterization of the D'Alembertian by the Poincar\'e Invariance \\[3mm] 
\normalsize \it{Dedicated to Professor Takayoshi Ogawa on the occasion of his sixtieth birthday}}

\author{Hiromichi Nakazato}
\email{hiromici@waseda.jp}
\affiliation{Department of Physics, Waseda University, Tokyo 169-8555, Japan}

\author{Tohru Ozawa}\thanks{Corresponding author.}
\email{txozawa@waseda.jp}
\affiliation{Department of Applied Physics, Waseda University, Tokyo 169-8555, Japan}


\begin{abstract}
\noindent
Many physical models are described by partial differential equations and the most important mathematical structure of the equations is governed by the corresponding linear partial differential operators. Those linear partial differential operators are sometimes determined by the symmetry under the group of motion. In this paper, the d'Alembertian is shown to be characterized as the only linear partial differential operator of the second order that is invariant under the Poincar\'e group and dilations in the Minkowski space-time $\mathbb R\times\mathbb R^n$. 
The method of proof depends on the analysis of the invariance of the corresponding polynomial in space-time under the time reflections and space rotations.
\end{abstract}


\maketitle

\section{Introduction}
\label{s:intro}
Many physical models are described by partial differential equations and the most important mathematical structure of the equations is governed by the corresponding linear partial differential operators. Those linear partial differential operators are sometimes determined by the symmetry under the group of motion. 

Symmetry is a decisive perspective in both mathematics and physics.  
In most cases, symmetry is recognized as invariance under specific operations such as permutations, translations, rotations, reflections, dilations and so on.  
Symmetry based on the invariance under transformations forms an 
essential concept in the Galois theory, crystallography, Roger Penrose's quasicrystals, Felix Klein's ``Erlangen Program" of geometry, the Cartan geometry as a natural generalization of ``Erlangen Program" \cite{Sharpe}, Emmy Noether's principle as well as the notion of gauge invariance introduced by Hermann Weyl in classical and quantum field theories \cite{FariaMelo,Scheck,Zeidler1,Zeidler2,Zeidler3}, and in the classification of elementary particles of the standard model \cite{FariaMelo,Zeidler1,Zeidler2,Zeidler3}.  

Among others, the Lorentz transformation plays a central role to support the concept of symmetry in relativistic physics.  
It is well known that Hendrik Lorentz introduced the transformation to explain the experimental result testifying the invariance of the speed of light, that Henri Poincar\'e proved the invariance of the Maxwell equations under the Lorentz transformation, and that Albert Einstein derived the Lorentz transformation from the principle of the invariance of the speed of light and of the special relativity \cite{Jackson,Scheck,Zeidler1,Zeidler2,Zeidler3}.  
These facts made a clear account of why natural laws in relativity must obey the Lorentz invariance.  

In this paper, we make a clear account of how the Lorentz invariance determines the structure of partial differential operators in space-time with constant coefficients.  
To be more specific, we study the symmetry of partial differential operators in space-time from the viewpoint of the invariance under the Lorentz transformations with translations and dilations.   
As a corollary, the d'Alemertian is shown to be characterized as the only second-order partial differential operator with constant coefficients that is invariant under Lorentz transformations and dilations.  
This reveals a special status of the d'Alembertian in the vector spaces consisting of all partial differential operators in space-time on the basis of the symmetry associated with the Poincar\'e group (generated by Lorentz transformations and translations) and dilations.

The proof of the main result depends on the reduction of the problem to the analysis of the corresponding polynomial in space-time, where the correspondence is given by exponential functions as in the theory of linear partial differential operators through the Fourier (or Laplace) transform \cite{Hormander}, thereby leading us to characterization problem of the polynomial by the Lorentz transformation.  
Then we make a separation of the space and time variables and show the invariance of the polynomial under time reflections and space rotations.  
Finally, we use the characterization of rotation-invariant polynomials in space variables.  

The paper is organized as follows.  
In Section~\ref{s:main}, we introduce basic notation in the Minkowski space-time and state our main results.  
In Section~\ref{s:proof}, we give proofs of the main results.  
Section~\ref{s:app} is the Appendix, where the characterization of rotation invariant polynomials as well as of the Laplacian is given in the Euclidean space. In Section \ref{s:final}, we discuss the corresponding problem in the quantum mechanics.

\section{Main Results}
\label{s:main}
In this section, we state our main results.  
To this end, we start introducing basic notation.  
We denote by $x=(t,\bm x)=(x^0,x^1,\ldots,x^n)$ a point in space-time $\mathbb R\times\mathbb R^n$.  
We also use the notation
\begin{equation*}
    x=\left[\begin{matrix}t\\\bm x\end{matrix}\right]=\left[\begin{matrix}x^0\\ x^1\\\vdots\\ x^n\end{matrix}\right]
\end{equation*}
as a column-vector representation to which $(1+n)\times(1+n)$-matrices apply.  
Let $(\bm e_0,\bm e_1,\ldots,\bm e_n)$ be the standard basis of $\mathbb R\times\mathbb R^n$, which is given by
\begin{equation}
    \bm e_0=\left[\begin{matrix}1\\0\\0\\\cdot\\\cdot\\\cdot\\0\end{matrix}\right]\begin{matrix}\lower2pt\hbox{$^{<0}$}\\ \\ \\ \\ \\ \\ \\\end{matrix}\;,\;
    \bm e_1=\left[\begin{matrix}0\\1\\0\\\cdot\\\cdot\\\cdot\\0\end{matrix}\right]\begin{matrix}\\\lower2pt\hbox{$^{<1}$}\\ \\ \\ \\ \\ \\\end{matrix}\;,\ldots,\;
    \bm e_j=\left[\begin{matrix}0\\\cdot\\0\\1\\0\\\cdot\\0\end{matrix}\right]\begin{matrix}\\ \\ \\\lower2pt\hbox{$^{<j}$} \\ \\ \\ \\\end{matrix}\;,\ldots,\;
    \bm e_n=\left[\begin{matrix}0\\0\\\cdot\\\cdot\\\cdot\\0\\1\end{matrix}\right]\begin{matrix}\\ \\ \\ \\ \\ \\\lower2pt\hbox{$^{<n}$}\end{matrix}
\end{equation}
as column vectors.  
Then the column vector representation of a point in $\mathbb R\times\mathbb R^n$ is given by 
\begin{equation}
    x=\left[\begin{matrix}t\\\bm x\end{matrix}\right]=x^0\bm e_0+x^j\bm e_j=x^\mu\bm e_\mu,
\end{equation}
where we have used Einstein's summation convention.  
Let $g=(g_{\mu\nu})$ be the standard metric tensor given by a $(1+n)\times(1+n)$-matrix as
\begin{equation}
    g=\left[\begin{matrix}1&&&&\\ &-1&&&\\&&\ddots&&\\&&&&-1\end{matrix}\right]=\hbox{\rm diag}(1,-1,\ldots,-1).
\end{equation}
Then the Minkowski metric $(\mathbb R\times\mathbb R^n)\times(\mathbb R\times\mathbb R^n)\to\mathbb R$ is given by
\begin{equation*}
    g_{\mu\nu}x^\mu y^\nu=ts-\bm x\cdot\bm y,
\end{equation*}
where $x=(t,\bm x)$, $y=(s,\bm y)$, and $\bm x\cdot\bm y=\sum\limits_{j=1}^nx^jy^j$ is the standard inner product in $\mathbb R^n$.  

Let ${\rm O}(n)$ be the orthogonal group defined by
\begin{equation}
    {\rm O}(n)=\{R\in{\rm GL}(n;\mathbb R); {}^t\!RR=R\,{}^t\!R=I\},
    \label{eq:orthogonal}
\end{equation}
where ${\rm GL}(n;\mathbb R)$ is the general linear group consisting of all invertible $n\times n$ matrices, $^t\!R$ the transposed matrix of $R$, and $I$ the identity matrix of order $n$.  
The condition on $R=(R^i\,{}_j)$ in (\ref{eq:orthogonal}) is written as
\begin{equation}
    R_k\,{}^iR^k\,{}_j=R^i\,{}_kR_j\,{}^k=\delta^i\,{}_j,
\end{equation}
where $\delta^i\,{}_j$ is Kronecker's delta.  
Let ${\rm O}(1,n)$ be the Lorentz group defined by
\begin{equation}
    {\rm O}(1,n)=\{\Lambda\in{\rm GL}(1+n;\mathbb R); {}^t\!\Lambda g\Lambda=\Lambda g{}^t\!\Lambda=g\},
\end{equation}
where the condition on $\Lambda=(\Lambda^\mu\,_\nu)$ is written as
\begin{equation}
    \Lambda^\alpha\,_\mu\, g_{\alpha\beta}\,\Lambda^\beta\,_\nu=\Lambda_\mu\,^\alpha\, g_{\alpha\beta}\,\Lambda_\nu\,^\beta=g_{\mu\nu}.
\end{equation}
The Laplacian $\triangle$ and d'Alembertian $\square$ are defined by
\begin{equation}
    \triangle=\nabla^2=\partial_1^2+\cdots\partial_n^2,\qquad
    \square=g^{\mu\nu}\partial_\mu\partial_\nu=\partial_0^2-\partial_1^2-\cdots-\partial_n^2=\partial_t^2-\triangle,
\end{equation}
where $\nabla=\bm\partial=(\partial_1,\ldots,\partial_n)$, $\partial_j=\partial/\partial x^j$, $\partial_0=\partial/\partial t$, and $(g^{\mu\nu})$ is the inverse of $(g_{\mu\nu})$, both of which are in fact ${\rm diag}(1,-1,\ldots,-1)$.

The action of $R\in{\rm O}(n)$ on $\mathbb R^n$ is understood as
\begin{equation}
    R\bm x=(R^j\,_kx^k)=\left[\begin{matrix}R^1\,_1&\cdots&R^1\,_n\\
    \vdots&&\vdots\\
    R^n\,_1&\cdots&R^n\,_n\end{matrix}\right]\left[\begin{matrix} x^1\\\vdots\\ x^n\end{matrix}\right]
\label{eq:Rx}
\end{equation}
and the action of $\Lambda\in{\rm O}(1,n)$ on $\mathbb R\times\mathbb R^n$ is understood as
\begin{equation}
    \Lambda x=(\Lambda^\mu\,_\nu x^\nu)=\Lambda\left[\begin{matrix}t\\\bm x\end{matrix}\right]=\left[\begin{matrix}\Lambda^0\,_0&\cdots&\Lambda^0\,_n\\
    \cdot&&\cdot\\
    \cdot&&\cdot\\
    \cdot&&\cdot\\
    \Lambda^n\,_0&\cdots&\Lambda^n\,_n\end{matrix}\right]\left[\begin{matrix}x^0\\x^1\\\cdot\\\cdot\\x^n\end{matrix}\right].
\end{equation}
With any $\bm y\in\mathbb R^n$, we associate the space translation $T_{\bm y}: \mathbb R^n\to\mathbb R^n$ by
\begin{equation}
    T_{\bm y}\,\bm x=\bm x-\bm y=(x^1-y^1,\ldots,x^n-y^n)=\left[\begin{matrix}x^1-y^1\\\vdots\\ x^n-y^n\end{matrix}\right]
\label{eq:Tyx}
\end{equation}
and with any $y=(s,\bm y)\in\mathbb R\times\mathbb R^n$, we associate the space-time translation $T_y: \mathbb R\times\mathbb R^n\to\mathbb R\times\mathbb R^n$ by
\begin{equation}
    T_y x=x-y=(t-s,\bm x-\bm y)=(x^0-y^0,x^1-y^1,\ldots,x^n-y^n)
    =\left[\begin{matrix}t-s\\\bm x-\bm y\end{matrix}\right]=\left[\begin{matrix}x^0-y^0\\x^1-y^1\\\vdots\\ x^n-y^n\end{matrix}\right].
\end{equation}
With $\lambda>0$, we associate dilation $D_\lambda: \mathbb R^n\to\mathbb R^n$ and $\mathbb R\times\mathbb R^n\to\mathbb R\times\mathbb R^n$ respectively by
\begin{equation}
    D_\lambda\bm x=\lambda\bm x\quad\hbox{and}\quad D_\lambda x=\lambda x=(\lambda t,\lambda\bm x).
\end{equation}
The Lorentz group, space-time translation, and dilation are defined to act on functions $u\in{\rm C}^\infty(\mathbb R\times\mathbb R^n;\mathbb C)$ as the corresponding pull-backs:
\begin{align}
    (\Lambda^*u)(x)&=u(\Lambda x)=u(\Lambda{\tiny\left[\begin{matrix}t\\\bm x\end{matrix}\right]}),\\
    (T_y^*u)(x)&=u(T_yx)=u(x-y)=u(t-s,\bm x-\bm y),\\
    (D_\lambda^*u)(x)&=u(D_\lambda x)=u(\lambda x)=u(\lambda t,\lambda\bm x).
\end{align}

We consider linear partial differential operators in $\mathbb R\times\mathbb R^n$ of order $m$ of the form
\begin{equation}
    L=\sum_{j+|\bm\alpha|\le m}a_{j\bm\alpha}\partial_t^j\bm\partial^{\bm\alpha},
\label{eq:L}
\end{equation}
where $\bm\alpha=(\alpha_1,\ldots,\alpha_n)\in\mathbb Z_{\ge0}^n$ is a multi-index with length $|\bm\alpha|=\alpha_1+\cdots+\alpha_n$, $\bm\partial^{\bm\alpha}=\partial_1^{\alpha_1}\cdots\partial_n^{\alpha_n}$, and $a_{j\bm\alpha}\in{\rm C}(\mathbb R\times\mathbb R^n;\mathbb C)$.  
We say that $L$ is \underline{\it Lorentz invariant\/} if and only if
\begin{equation}
    \Lambda^*Lu=L\Lambda^*u
\end{equation}
for any $\Lambda\in{\rm O}(1,n)$ and any $u\in{\rm C}^\infty(\mathbb R\times\mathbb R^n;\,\mathbb C)$.  
We say that $L$ is \underline{\it Poincar\'e invariant\/} if and only if $L$ is Lorentz invariant and \underline{\it translation invariant\/}, namely,
\begin{equation*}
    T_y^*Lu=LT_y^*u
\end{equation*}
for any $y\in\mathbb R\times\mathbb R^n$ and any $u\in{\rm C}^\infty(\mathbb R\times\mathbb R^n;\mathbb C)$.  
We say that $L$ is \underline{\it dilation invariant\/} if and only if for any $u\in{\rm C}^\infty(\mathbb R\times\mathbb R^n;\mathbb C)\backslash\{0\}$, the following statements are equivalent:
\begin{itemize}
\item[(i)] $Lu=0$.
\item[(ii)] $LD_\lambda^*u=0$ for any $\lambda>0$.
\item[(iii)] $LD_\lambda^*u=0$ for some $\lambda\not=1$.
\end{itemize}

We now state our main results.
\begin{theorem} \label{T:1}
For any partial differential operator in space-time $\mathbb R\times\mathbb R^n$ of order $m$ of the form    
\begin{equation*}
    L=\sum_{j+|\bm\alpha|\le m}a_{j\bm\alpha}\partial_t^j\bm\partial^{\bm\alpha},
\end{equation*}
with $a_{j\bm\alpha}\in{\rm C}(\mathbb R\times\mathbb R^n;\mathbb C)$ and $\sum\limits_{j+|\bm\alpha|=m}|a_{j\bm\alpha}|\not=0$, the following three statements are equivalent.
\begin{itemize}
    \item[(1)] $L$ is Poincar\'e invariant.  
    Namely,
    \begin{itemize}
        \item[(i)] For any $y\in\mathbb R\times\mathbb R^n$ and any $u\in{\rm C}^\infty(\mathbb R\times\mathbb R^n;\mathbb C)$, $T_y^*Lu=LT_y^*u$.
        \item[(ii)] For any $\Lambda\in{\rm O}(1,n)$ and any $u\in{\rm C}^\infty(\mathbb R\times\mathbb R^n;\mathbb C)$, $\Lambda^*Lu=L\Lambda^*u$.
    \end{itemize}
    \item[(2)] All coefficients of $L$ are constants and $L$ is Lorentz invariant.  
    Namely,
    \begin{itemize}
    \item[(i)] For any $(j,\bm\alpha)\in\mathbb Z_{\ge0}\times\mathbb Z_{\ge0}^n$ with $j+|\bm\alpha|\le m$ and any $x\in\mathbb R\times\mathbb R^n$, $a_{j\bm\alpha}(x)=a_{j\bm\alpha}(0)$.
    \item[(ii)] For any $\Lambda\in{\rm O}(1,n)$ and any $u\in{\rm C}^\infty(\mathbb R\times\mathbb R^n;\mathbb C)$, $\Lambda^*Lu=L\Lambda^*u$.
    \end{itemize}
    \item[(3)]  $L$ is given by a polynomial in $\square$ of order $[m/2]\coloneqq{\rm max}\{k\in\mathbb Z_{\ge0}; k\le m/2\}$.  
    Namely, there exists $(b_j;0\le j\le[m/2])\subset\mathbb C$ such that $b_{[m/2]}\not=0$ and $L=\sum\limits_{j=0}^{[m/2]}b_j\square^j$.
\end{itemize}
\end{theorem}

\begin{corollary} \label{c:1}
    Any Poincar\'e invariant partial differential operator of the second order in space-time $\mathbb R\times\mathbb R^n$ is represented as $L=\alpha\square+\beta$ with $\alpha\in\mathbb C\backslash\{0\}$ and $\beta\in\mathbb C$.
\end{corollary}

\begin{corollary} \label{c:2}
    Any Poincar\'e and dilation invariant partial differential operator of the second order in space-time $\mathbb R\times\mathbb R^n$ is represented as $L=\alpha\square$ with $\alpha\in\mathbb C\backslash\{0\}$.
\end{corollary}

\begin{remark} \label{r:1}
    The Poincar\'e invariance of the d'Alembertian, namely, the implication $(3)\Rightarrow(1)$ is well-known \cite{FariaMelo,Jackson,Scheck,Zeidler1,Zeidler2,Zeidler3}.  
    The main purpose of this paper is to prove the converse implication $(1)\Rightarrow(3)$, thereby giving a characterization of the d'Alembertian in terms of the Poincar\'e or Lorentz invariance.
\end{remark}

\begin{remark} \label{r:2}
    The definition of dilation invariance in this paper is weaker than the invariance given by the commutation relation $D_\lambda^*L=LD_\lambda^*$, which is unnatural for dilations since $\square D_\lambda^*=\lambda^2D_\lambda^*\square$ with additional factor $\lambda^2$ arising in the second derivatives.
\end{remark}

\begin{remark} \label{r:3}
    We consider linear partial differential operators $L$ of the specific form (\ref{eq:L}).  
    It is not a restrictive assumption in the framework of all linear operators in ${\rm C}^\infty(\mathbb R\times\mathbb R^n;\mathbb C)$ in the sense that any linear operator from ${\rm C}^\infty_0(\mathbb R\times\mathbb R^n;\mathbb C)$ to ${\rm C}^\infty(\mathbb R\times\mathbb R^n;\mathbb C)$ with the local property:
\begin{equation*}
    {\rm supp}(Lu)\subset{\rm supp}\,u\quad \hbox{for all}\quad u\in{\rm C}^\infty_0(\mathbb R\times\mathbb R^n;\mathbb C)
\end{equation*}
    is uniquely represented as (\ref{eq:L}) at least locally \cite{Peetre}, where {\rm supp} denotes the support of functions and ${\rm C}^\infty_0(\mathbb R\times\mathbb R^n;\mathbb C)$ is the vector space consisting of all compactly supported smooth functions.
\end{remark}
    
Theorem~\ref{T:1} shows that the Poincar\'e invariance determines the structure of the space of partial differential operators that is exclusively generated by the d'Alembertian and that the Lorentz invariance determines the structure of the space of partial differential operators with constant coefficients that is exclusively generated by the d'Alembertian as well.  
Conversely, Corollary~\ref{c:1} characterizes the d'Alembertian up to mass term in terms of the Poincar\'e [resp.~Lorentz] invariance of partial differential operators [resp.~constant coefficients] on space-time.  
Corollary~\ref{c:2} shows that the mass term can be deleted away under the dilation invariance.  

We prove the main results in the next section.  
The first step in the proof is to reduce the problem for $L$ to that of the polynomial in $\xi$ via
\begin{equation}
    (Le_\xi)(x)=p(x,\xi)e_\xi(x),
    \label{eq:Lexi}
\end{equation}
where $e_\xi:\mathbb R\times\mathbb R^n\ni x=(t,\bm x)\mapsto e_\xi(x)=\exp(\tau t+\bm\xi\cdot\bm x)\in\mathbb R$,
\begin{equation}
    p(x,\xi)=\sum_{j+|\bm\alpha|\le m}a_{j\bm\alpha}(x)\tau^j\bm\xi^{\bm\alpha}
\end{equation}
for $\xi=(\tau,\bm\xi)\in\mathbb R\times\mathbb R^n$, where $\bm\xi^{\bm\alpha}=\xi_1^{\alpha_1}\cdots\xi_n^{\alpha_n}$.  
By (\ref{eq:Lexi}), the symmetry of $L$ is described by the symmetry of $p$.  
By the translation invariance of $L$, we show that $p(x,\xi)=p(0,\xi)$, namely, $p$ is a polynomial in $\xi$ with constant coefficients $(a_{j\bm\alpha}(0); j+|\bm\alpha|\le m)$: 
\begin{equation}
    p(\xi)\coloneqq p(0,\xi)=\sum_{j+|\bm\alpha|\le m}a_{j\bm\alpha}(0)\tau^j\bm\xi^{\bm\alpha}.
\label{eq:pxi}
\end{equation}
By (\ref{eq:Lexi}) and (\ref{eq:pxi}), we obtain
\begin{equation}
    Le_\xi=p(\xi)e_\xi,
\end{equation}
which shows that  $p(\xi)$ is an eigenvalue of $L$ with eigenfunction $e_\xi$.  
Then we separate the time and space variables $(\tau,\bm\xi)$ in (\ref{eq:pxi}) and show the invariance of $p(\xi)$ under time reflection and space rotation.  
Finally, we use the characterization of rotation invariant polynomial in $\bm\xi\in\mathbb R^n$.

\section{Proof of the Main Results}
\label{s:proof}
In this section, we prove Theorem~\ref{T:1} and Corollary~\ref{c:2}, since Corollary~\ref{c:1} follows directly from Theorem~\ref{T:1} with $m=2$.  

\medskip
\noindent\underline{\it Proof of Theorem~\ref{T:1}\/}

\medskip
For $\xi=(\tau,\bm\xi)\in\mathbb R\times\mathbb R^n$, we introduce $e_\xi\in{\rm C}^\infty(\mathbb R\times\mathbb R^n;\mathbb R)$ by
\begin{equation}
    e_\xi(x)=e_\xi(t,\bm x)=\exp(\tau t+\bm\xi\cdot\bm x)\quad\hbox{\rm for}\quad x=(t,\bm x)\in\mathbb R\times\mathbb R^n.
\label{eq:exi}
\end{equation}
Then we see that $e_\xi(0)=1$ and for any $(j,\bm\alpha)\in\mathbb Z_{\ge0}\times\mathbb Z^n_{\ge0}$,
\begin{equation}
    \partial_t^j\bm\partial^{\bm\alpha}e_\xi=\tau^i\bm\xi^{\bm\alpha}e_\xi.
\label{eq:partial}
\end{equation}
For $x,\xi\in\mathbb R\times\mathbb R^n$, we define
\begin{equation}
    p(x,\xi)=\sum_{j+|\bm\alpha|\le m}a_{j\bm\alpha}(x)\tau^j\bm\xi^{\bm\alpha}.
\label{eq:pxxi}
\end{equation}
By (\ref{eq:exi}), (\ref{eq:partial}), and (\ref{eq:pxxi}), we have
\begin{equation}
    (Le_\xi)(x)=p(x,\xi)e_\xi(x)
\label{eq:Lexi2}
\end{equation}
for any $x,\xi\in\mathbb R\times\mathbb R^n$.  
We use the equality (\ref{eq:Lexi2}) in the subsequent argument.

\medskip
\noindent$(1)\Rightarrow(2)$: It suffices to prove (1)(i)$\Rightarrow$(2)(i).  
The translation invariance described in (1)(i) with $u=e_\xi$ implies
\begin{equation}
    T_y^*Le_\xi=LT_y^*e_\xi\quad\hbox{\rm for any}\quad y=(s,\bm y)\in\mathbb R\times\mathbb R^n.
\label{eq:TLexi}
\end{equation}
The LHS of (\ref{eq:TLexi}) at $x=(t,\bm x)$ is given by
\begin{equation}
    (T_y^*Le_\xi)(x)=(Le_\xi)(x-y)=p(x-y,\xi)e_\xi(x-y),
    \label{eq:TLexi2}
\end{equation}
where we have used (\ref{eq:Lexi2}), while the RHS of (\ref{eq:TLexi}) is given by
\begin{align}
    (LT_y^*e_\xi)(x)&=\sum_{j+|\bm\alpha|\le m}a_{j\bm\alpha}(x)(\partial_t^j\bm\partial^{\bm\alpha}T_y^*e_\xi)(x)=\sum_{j+|\bm\alpha|\le m}a_{j\bm\alpha}(x)(T_y^*\partial_t^j\bm\partial^{\bm\alpha}e_\xi)(x)\nonumber\\
    &=\sum_{j+|\bm\alpha|\le m}a_{j\bm\alpha}(x)\tau^j\bm\xi^{\bm\alpha}e_\xi(x-y)=p(x,\xi)e_\xi(x-y),
\label{eq:LTexi}
\end{align}
where we have used (\ref{eq:partial}) and (\ref{eq:pxxi}).  
By (\ref{eq:TLexi}), (\ref{eq:TLexi2}), and (\ref{eq:LTexi}), with $y=x$, we have
\begin{equation}
    p(0,\xi)=p(x,\xi)
\label{eq:p0xi}
\end{equation}
since $e_\xi(0)=1$.  
Since (\ref{eq:p0xi}) holds for any $\xi\in\mathbb R\times\mathbb R^n$, we have
\begin{equation*}
    a_{j\bm\alpha}(x)=a_{j\bm\alpha}(0)
\end{equation*}
for any $x\in\mathbb R\times\mathbb R^n$ and any $(j,\bm\alpha)\in\mathbb Z_{\ge0}\times\mathbb Z_{\ge0}^n$ with $j+|\bm\alpha|\le m$, as required.

\medskip
\noindent$(2)\Rightarrow(3)$: We first note that
\begin{equation*}
    (\Lambda^*e_\xi)(x)=e_\xi(\Lambda x)=e_{\tiny\left[\begin{matrix}\tau\\\bm\xi\end{matrix}\right]}(\Lambda{\tiny\left[\begin{matrix}t\\\bm x\end{matrix}\right]})=\exp({\tiny\left[\begin{matrix}\tau\\\bm\xi\end{matrix}\right]}\cdot\Lambda{\tiny\left[\begin{matrix}t\\\bm x\end{matrix}\right]})=\exp(({}^t\!\Lambda{\tiny\left[\begin{matrix}\tau\\\bm\xi\end{matrix}\right]})\cdot{\tiny\left[\begin{matrix}t\\\bm x\end{matrix}\right]})=e_{{}^t\!\Lambda\xi}(x),
\end{equation*}
namely,
\begin{equation}
    \Lambda^*e_\xi=e_{^t\!\Lambda\xi}.
\label{eq:Lambdaexi}
\end{equation}
We write $\Lambda\in{\rm O}(1,n)$ in the form
\begin{equation*}
    \Lambda=\left[\begin{matrix}s&u_1&\cdots&u_n\\v_1&&&\\\cdot&&&\\\cdot&&R&\\\cdot&&&&\\ v_n&&&&\end{matrix}\right]
    =\left[\begin{matrix}s&^t\!\bm u\\\bm v&R\end{matrix}\right]\quad{\rm with}\quad\bm v=\left[\begin{matrix}v_1\\\cdot\\\cdot\\\cdot\\ v_n\end{matrix}\right],\quad ^t\!\bm u=[u_1\cdots u_n],
\end{equation*}
and $n\times n$ matrix $R$.  
We have
\begin{equation}
    ^t\!\Lambda\xi=\left[\begin{matrix}s&^t\!\bm v\\\bm u&^t\!R\end{matrix}\right]\left[\begin{matrix}\tau\\\bm\xi\end{matrix}\right]=\left[\begin{matrix}s\tau+\bm v\cdot\bm\xi\\\tau\bm u+{}^t\!R\bm\xi\end{matrix}\right],
\label{eq:tLambdaxi}
\end{equation}
\begin{equation}
    \partial_t^j\bm\partial^{\bm\alpha}\Lambda^*e_\xi=\partial_t^j\bm\partial^{\bm\alpha}e_{^t\!\Lambda\xi}=(s\tau+\bm v\cdot\bm\xi)^j(\tau\bm u+{}^t\!R\bm\xi)^{\bm\alpha}e_{^t\!\Lambda\xi}.
\label{eq:deldelLexi}
\end{equation}
By (2)(ii), we have
\begin{equation}
    \Lambda^*Le_\xi=L\Lambda^*e_\xi,
\label{eq:LLe}
\end{equation}
where the LHS of (\ref{eq:LLe}) is given by
\begin{equation}
    \Lambda^*Le_\xi=\Lambda^*(p(0,\xi)e_\xi)=p(0,\xi)\Lambda^*e_\xi=p(0,\xi)e_{^t\!\Lambda\xi},
\label{eq:LambdaLexi}
\end{equation}
where we have used (\ref{eq:Lexi2}), (\ref{eq:p0xi}), and (\ref{eq:Lambdaexi}), while the RHS of (\ref{eq:LLe}) is given by 
\begin{equation}
    L\Lambda^*e_\xi=\sum_{j+|\bm\alpha|\le m}a_{j\bm\alpha}(0)(s\tau+\bm v\cdot\bm\xi)^j(\tau\bm u+{}^t\!R\bm\xi)^{\bm\alpha}e_{^t\!\Lambda\xi},
\label{eq:LLexi}
\end{equation}
where we have used (\ref{eq:deldelLexi}).  
We define the following polynomial $p$ in $\xi$ by
\begin{equation}
    p(\xi)=p(0,\xi)=\sum_{j+|\bm\alpha|\le m}a_{j\bm\alpha}(0)\tau^j\bm\xi^{\bm\alpha}\quad{\rm for}\quad\xi=(\tau,\bm\xi)\in\mathbb R\times\mathbb R^n.
\end{equation}
Then we have
\begin{equation}
    (^t\!\Lambda^*p)(\xi)=p({}^t\!\Lambda\xi)=p(s\tau+\bm v\cdot\bm\xi,\tau\bm u+{}^t\!R\bm\xi)=(L\Lambda^*e_\xi)(0)=(\Lambda^*Le_\xi)(0)=p(\xi),
\label{eq:tLambdap}
\end{equation}
where we have used (\ref{eq:tLambdaxi}), (\ref{eq:LLexi}), (\ref{eq:LLe}), and (\ref{eq:LambdaLexi}).
By (\ref{eq:tLambdap}), 
\begin{equation*}
    ^t\!\Lambda^*p=p
\end{equation*}
for any $\Lambda\in{\rm O}(1,n)$.  
Since $^t\!\Lambda\in{\rm O}(1,n)$, changing $\Lambda\mapsto{}^t\!\Lambda$,  we see that 
\begin{equation}
    \Lambda^*p=p
\label{eq:Lambdap}
\end{equation}
for any $\Lambda\in{\rm O}(1,n)$.  
In particular, setting $s=1,\,\bm u=\bm v=\bm0$, and $R\in{\rm O}(n)$, the corresponding $\Lambda$ given by $\Lambda=1\otimes R$ belongs to ${\rm O}(1,n)$.  
For $j\in\{0,1,\ldots,m\}$, we define the following polynomial $p_j$ in $\bm\xi$ of order at most $m-j$ by
\begin{equation}
    p_j(\bm\xi)=\sum_{|\bm\alpha|\le m-j}a_{j\bm\alpha}(0)\bm\xi^{\bm\alpha}.
\label{eq:pxi0}
\end{equation}
Then $p$ is represented as
\begin{equation}
    p(\xi)=\sum_{j=0}^mp_j(\bm\xi)\tau^j
\label{eq:pxi1}
\end{equation}
and the invariance of $p$ under $\Lambda$ (\ref{eq:Lambdap}) is rewritten with $\Lambda=1\otimes R$ as
\begin{equation}
    p(\xi)=((1\otimes R)^*p)(\xi)=\sum_{j=0}^m(R^*p_j)(\bm\xi)\tau^j
\label{eq:pxi2}
\end{equation}
for any $\xi=(\tau,\bm\xi)\in\mathbb R\times\mathbb R^n$.  
It follows from (\ref{eq:pxi1}) and (\ref{eq:pxi2}) that
\begin{equation}
    R^*p_j=p_j
\label{eq:Rpj}
\end{equation}
for any $R\in{\rm O}(n)$ and any $j\in\{0,1,\ldots,m\}$.

By Proposition~\ref{p:1} in Section~\ref{s:app}, (\ref{eq:Rpj}) implies that $p_j$ is written as a polynomial in $|\bm\xi|^2=\sum\limits_{j=1}^n\xi_j^2$.  
(Proposition~\ref{p:1} is for homogeneous polynomials, while $p_j$ given by (\ref{eq:pxi0}) is inhomogeneous. The statement actually follows from Proposition~\ref{p:1} by means of the argument from (\ref{eq:p2-a}) to (\ref{eq:Rpj-a}) below.) 
To be more specific, for any $j\in\{0,1,\ldots,m\}$ there exists a finite sequence $(b_{jk};0\le k\le[(m-j)/2])\subset\mathbb C$ such that
\begin{equation}
    p_j(\bm\xi)=\sum_{k=0}^{[(m-j)/2]}b_{jk}|\bm\xi|^{2k}
\end{equation}
for any $\bm\xi\in\mathbb R^n$.  
For $\ell\in\{0,1,\ldots,m\}$, we introduce the following homogeneous polynomial $p_\ell$ in $\xi=(\tau,\bm\xi)$ of order $\ell$ as
\begin{equation}
    p_\ell(\xi)=p_\ell(\tau,\bm\xi)=\sum_{j+2k=\ell}b_{jk}\tau^j|\bm\xi|^{2k}.
\label{eq:pellxi}
\end{equation}
Then $p$ is rewritten as
\begin{equation}
    p(\xi)=\sum_{j=0}^mp_j(\bm\xi)\tau^j=\sum_{j=0}^m\Bigl(\sum_{k=0}^{[(m-j)/2]}b_{jk}|\bm\xi|^{2k}\Bigr)\tau^j=\sum_{\ell=0}^m\bigl(\sum_{j+2k=\ell}b_{jk}\tau^j|\bm\xi|^{2k}\bigr)=\sum_{\ell=0}^mp_\ell(\xi).
\label{eq:pxi3}
\end{equation}
By (\ref{eq:Lambdap}), (\ref{eq:pxi3}), and the homogeneity of degree $\ell$ of $p_\ell$ in (\ref{eq:pellxi}), we have
\begin{equation}
    \sum_{\ell=0}^m(\Lambda^*p_\ell)(\xi)r^\ell=\sum_{\ell=0}^m(\Lambda^*p_\ell)(r\xi)=(\Lambda^*p)(r\xi)=p(r\xi)=\sum_{\ell=0}^mp_\ell(r\xi)=\sum_{\ell=0}^mp_\ell(\xi)r^\ell
\end{equation}
for any $\xi\in\mathbb R\times\mathbb R^n$ and any $r>0$.  
This implies that
\begin{equation}
    \Lambda^*p_\ell=p_\ell
\label{eq:Lambdapell}
\end{equation}
for any $\Lambda\in{\rm O}(1,n)$ and any $\ell\in\{0,1,\ldots,m\}$.  

We apply $\Lambda=-I\in{\rm O}(1,n)$ to (\ref{eq:Lambdapell}) to obtain
\begin{equation}
    p_\ell(-\tau,\bm\xi)=p_\ell(-\tau,-\bm\xi)=p_\ell(\tau,\bm\xi).
\label{eq:pell-}
\end{equation}
For any $\ell\in\{0,1,\ldots, m\}$, $\tau\in\mathbb R$, and $r>0$, we set $\bm\xi=\sqrt{r}\bm e_1$ and apply (\ref{eq:pell-}) to obtain
\begin{align}
    \sum_{k=0}^{[\ell/2]}\bigl((-1)^{\ell-2k}b_{\ell-2k,k}\tau^{\ell-2k}\bigr)r^k&=\sum_{j+2k=\ell}(-1)^jb_{jk}\tau^jr^k=p_\ell(-\tau,\sqrt{r}\bm e_1)\nonumber\\
    &=p_\ell(\tau,\sqrt{r}\bm e_1)=\sum_{j+2k=\ell}b_{jk}\tau^jr^k=\sum_{k=0}^{[\ell/2]}\bigl(b_{\ell-2k,k}\tau^{\ell-2k}\bigr)r^k.
\label{eq:pell2}
\end{align}
By (\ref{eq:pell2}), we see that
\begin{equation}
    (-1)^{\ell-2k}b_{\ell-2k,k}=b_{\ell-2k,k}
\label{eq:b}
\end{equation}
for any $\ell,k$ with $0\le k\le[\ell/2]$ and $0\le\ell\le m$. 
It follows from (\ref{eq:b}) that $\ell$ is even or $b_{\ell-2k,k}=0$ and, equivalently, that $j=\ell-2k$ is even or $b_{jk}=0$.  
This shows that $p_\ell$ takes the form 
\begin{equation}
    p_\ell(\tau,\bm\xi)=\begin{cases}\displaystyle\sum_{2j+2k=\ell}b_{2j,k}\tau^{2j}|\bm\xi|^{2k}\;&\hbox{if $\ell$ is even,}\\\quad0\;&\hbox{if $\ell$ is odd.}\end{cases}
\label{eq:pell3}
\end{equation}

Let $\Gamma$ be the light cone defined by $\Gamma=\{\xi\in\mathbb R\times\mathbb R^n; g_{\mu\nu}\xi^\mu\xi^\nu=0\}=\{(\tau,\bm\xi)\in\mathbb R\times\mathbb R^n; \tau^2=|\bm\xi|^2\}$.  
For $\ell\in\{0,1,\ldots,m\}$, we define $\varphi_\ell: (\mathbb R\times\mathbb R^n)\backslash\Gamma\mapsto\mathbb R$ by
\begin{equation}
    \varphi_\ell(\tau,\bm\xi)={p_\ell(\tau,\bm\xi)\over\bigl|\tau^2-|\bm\xi|^2\bigr|^{\ell/2}},\quad(\tau,\bm\xi)\in(\mathbb R\times\mathbb R^n)\backslash\Gamma,
\end{equation}
where $\varphi_0$ is understood as $\varphi_0(\tau,\bm\xi)=p_0(\tau,\bm\xi)=b_{00}$.  

We prove that $\varphi_\ell$ is a locally constant function on $(\mathbb R\times\mathbb R^n)\backslash\Gamma$.  
It suffices to prove that $\varphi_\ell$ is a locally constant function on $((0,\infty)\times\mathbb R^n)\backslash\Gamma$ since $\varphi_\ell(-\tau,\bm\xi)=\varphi_\ell (\tau,\bm\xi)$.  
We notice the following basic properties:
\begin{itemize}
    \item (positive homogeneity of order 0)
    \item[] For any $r>0$ and any $(\tau,\bm\xi)\in((0,\infty)\times\mathbb R^n)\backslash\Gamma$, 
\begin{equation}
\varphi_\ell(r\tau,r\bm\xi)=\varphi_\ell(\tau,\bm\xi).
\label{eq:order0}
\end{equation}
    \item (Lorentz invariance)
    \item[] For any $\Lambda\in{\rm O}(1,n)$ and any $(\tau,\bm\xi)\in((0,\infty)\times\mathbb R^n)\backslash\Gamma$,
\begin{equation}
    (\Lambda^*\varphi_\ell)(\tau,\bm\xi)=\varphi_\ell(\tau,\bm\xi).
\label{eq:Lorentz}
\end{equation}
\end{itemize}

From now on we distinguish two cases: (i) $\tau>|\bm\xi|\geq0$, (ii) $|\bm\xi|>\tau>0$.  

\medskip
\underline{Case (i)}: There exists $\Lambda\in{\rm O}(1,n)$ such that $\Lambda\xi=(g_{\mu\nu}\xi^\mu\xi^\nu)^{1/2}\bm e_0=(\tau^2-|\bm\xi|^2)^{1/2}\bm e_0$.  
By (\ref{eq:order0}) and (\ref{eq:Lorentz}), we have
\begin{equation*}
    \varphi_\ell(\xi)=(\Lambda^*\varphi_\ell)(\xi)=\varphi_\ell(\Lambda\xi)=\varphi_\ell\bigl((\tau^2-|\bm\xi|^2)^{1\over2}\bm e_0\bigr)=\varphi_\ell(\bm e_0)=p_\ell(\bm e_0)=b_{\ell0}.
\end{equation*}
This implies that 
\begin{equation}
    p_\ell(\tau,\bm\xi)=b_{\ell0}(\tau^2-|\bm\xi|^2)^{\ell\over2}.
\label{eq:pell4}
\end{equation}
Moreover, by (\ref{eq:pell3}) and (\ref{eq:pell4}), we have
\begin{equation}
    \sum_{2j+2k=\ell}b_{2j,k}\tau^{2j}|\bm\xi|^{2k}=p_\ell(\tau,\bm\xi)=b_{\ell0}(\tau^2-|\bm\xi|^2)^{\ell\over2}=b_{\ell0}\sum_{i=0}^{\ell/2}\begin{pmatrix}\ell/2\\ i\end{pmatrix}\tau^{\ell-2i}(-|\bm\xi|^2)^i.
\label{eq:pell5}
\end{equation}
Comparing both terms in (\ref{eq:pell5}) with $(j,k)=(0,\ell/2)$ on the LHS and $i=\ell/2$ on the RHS, we obtain
\begin{equation}
    b_{0{\ell\over2}}=(-1)^{{\ell\over2}}b_{\ell0}.
\label{eq:b0ell}
\end{equation}

\underline{Case (ii)}: There exists $\Lambda\in{\rm O}(1,n)$ such that $\Lambda\xi=(-g_{\mu\nu}\xi^\mu\xi^\nu)^{1/2}\bm e_1=(|\bm\xi|^2-\tau^2)^{1\over2}\bm e_1$.  
By (\ref{eq:order0}) and (\ref{eq:Lorentz}), we have
\begin{equation}
    \varphi_\ell(\xi)=(\Lambda^*\varphi_\ell)(\xi)=\varphi_\ell(\Lambda\xi)=\varphi_\ell\bigl((|\bm\xi|^2-\tau^2)^{1\over2}\bm e_1\bigr)=\varphi_\ell(\bm e_1)=p_\ell(\bm e_1)=b_{0{\ell\over2}}.
\label{eq:phiellxi}
\end{equation}
By (\ref{eq:b0ell}) and (\ref{eq:phiellxi}), we obtain
\begin{equation}
    p_\ell(\tau,\bm\xi)=b_{0{\ell\over2}}(|\bm\xi|^2-\tau^2)^{\ell\over2}=b_{\ell0}(\tau^2-|\bm\xi|^2)^{\ell\over2}.
\end{equation}

In both cases, we have proved that $\varphi_\ell$ is a locally constant function and that 
\begin{equation}
    p_\ell(\tau,\bm\xi)=b_{\ell0}(\tau^2-|\bm\xi|^2)^{\ell\over2}
\label{eq:pell6}
\end{equation}
for any $(\tau,\bm\xi)\in(\mathbb R\times\mathbb R^n)\backslash\Gamma$.  
If $(\tau,\bm\xi)\in\Gamma$, the RHS of (\ref{eq:pell6}) is zero, while the LHS is given by a polynomial, a continuous function on $\mathbb R\times\mathbb R^n$.  
This proves that (\ref{eq:pell6}) holds for any $(\tau,\bm\xi)\in\mathbb R\times\mathbb R^n$.  
Therefore,
\begin{equation}
    p(\tau,\bm\xi)=\sum_{\ell=0}^mp_\ell(\tau,\bm\xi)=\sum_{j=0}^{[m/2]}p_{2j}(\tau,\bm\xi)=\sum_{j=0}^{[m/2]}b_{2j,0}(\tau^2-|\bm\xi|^2)^j.
\end{equation}
This yields
\begin{equation*}
    L=\sum_{j=0}^{[m/2]}b_j\square^j
\end{equation*}
if we define $b_j=b_{2j,0}$.  
We notice that $b_{[m/2]}\not=0$ since the coefficients of the highest order $m$ of $L$ do not vanish identically.

\medskip
\noindent$(3)\Rightarrow(1)$: It suffices to prove that the Poincar\'e invariance of the d'Alembertian $\square$.  
The commutativity between space-time derivatives and translations implies the invariance of $\square$ under space-time translations.  
For $\Lambda\in{\rm O}(1,n)$, we write $\Lambda=(\Lambda^\nu\,_\mu)$.  
By the differentiation of composite functions, we have 
\begin{equation*}
    (\partial_\mu\Lambda^*u)(x)=\partial_\mu(u(\Lambda x))=\partial_\nu u(\Lambda x)\partial_\mu(\Lambda x)^\nu=(\Lambda^*\partial_\nu u)(x)\Lambda^\nu\,_\mu
\end{equation*}
so that
\begin{equation}
    \partial_\mu\Lambda^*=\Lambda^\nu\,_\mu\Lambda^*\partial_\nu,
\label{eq:LL}
\end{equation}
which in turn implies
\begin{align*}
    \square\Lambda^*&=g^{\lambda\mu}\partial_\lambda\partial_\mu\Lambda^*=g^{\lambda\mu}\partial_\lambda(\Lambda^\nu\,_\mu\Lambda^*\partial_\nu)=g^{\lambda\mu}\Lambda^\nu\,_\mu(\partial_\lambda\Lambda^*)\partial_\nu\\
    &=g^{\lambda\mu}\Lambda^\nu\,_\mu(\Lambda^\rho\,_\lambda\Lambda^*\partial_\rho)\partial_\nu=(\Lambda^\rho\,_\lambda g^{\lambda\mu}\Lambda^\nu\,_\mu)\Lambda^*\partial_\rho\partial_\nu=g^{\rho\nu}\Lambda^*\partial_\rho\partial_\nu=\Lambda^*g^{\rho\nu}\partial_\rho\partial_\nu\\
    &=\Lambda^*\square
\end{align*}
as required.  
This completes the proof of Theorem~\ref{T:1}.

\medskip
\noindent\underline{\it Proof of Corollary~\ref{c:2}\/}

\smallskip
It suffices to prove that the dilation invariance of $\square+\gamma$ implies $\gamma=0$.  
We assume that $\square+\gamma$ is dilation invariant and that $u\in{\rm C}^\infty(\mathbb R\times\mathbb R^n;\mathbb C)\backslash\{0\}$ satisfies $\square u+\gamma u=0$.  
Then $(\square+\gamma)D_\lambda^*u=0$ for some $\lambda\not=1$.  
This means
\begin{equation}
    \lambda^2(\square u)(\lambda x)+\gamma u(\lambda x)=0
\label{eq:dil1}
\end{equation}
for any $x\in\mathbb R\times\mathbb R^n$, while solution $u$ of $\square u+\gamma u=0$ satisfies
\begin{equation}
    (\square u)(\lambda x)+\gamma u(\lambda x)=0
\label{eq:dil2}
\end{equation}
for any $x\in\mathbb R\times\mathbb R^n$.  
By (\ref{eq:dil1}) and (\ref{eq:dil2}), we obtain
\begin{equation}
    (\lambda^2-1)\gamma u(\lambda x)=0
\label{eq:dil3}
\end{equation}
for any $x\in\mathbb R\times\mathbb R^n$.  
Since $u\not=0$, there exists $x_0\in\mathbb R\times\mathbb R^n$ such that $u(x_0)\not=0$.  
We apply (\ref{eq:dil3}) with $x=\lambda^{-1}x_0$ to obtain
\begin{equation*}
    (\lambda^2-1)\gamma u(x_0)=0,
\end{equation*}
which concludes $\gamma=0$, as required.

\section{Appendix: Characterization of the Laplacian by the Euclidean motion group}
\label{s:app}

In this section, we review the characterization of the Laplacian by the Euclidean motion group \cite{Nomura,Shimakura}.  
We designate a point in the Euclid space $\mathbb R^n$ as
\begin{equation*}
    \bm x=(x^1,\ldots,x^n)=\left[\begin{matrix}x^1\\\cdot\\\cdot\\\cdot\\ x^n\end{matrix}\right]=x^j\bm e_j,
\end{equation*}
where $(\bm e_j;j=1,\ldots,n)$ is the standard basis of $\mathbb R^n$ given by
\begin{equation*}
    \bm e_1=\left[\begin{matrix}1\\0\\0\\\cdot\\\cdot\\\cdot\\0\end{matrix}\right]\begin{matrix}\lower2pt\hbox{$^{<1}$}\\ \\ \\ \\ \\ \\ \\\end{matrix}\;,\;
    \bm e_j=\left[\begin{matrix}0\\\cdot\\0\\1\\0\\\cdot\\0\end{matrix}\right]\begin{matrix}\\ \\ \\\lower2pt\hbox{$^{<j}$} \\ \\ \\ \\\end{matrix}\;,\ldots,\;
    \bm e_n=\left[\begin{matrix}0\\0\\\cdot\\\cdot\\\cdot\\0\\1\end{matrix}\right]\begin{matrix}\\ \\ \\ \\ \\ \\\lower2pt\hbox{$^{<n}$}\end{matrix}.
\end{equation*}
For $\bm x=(x^1,\ldots,x^n)$ and $\bm y=(y^1,\ldots,y^n)$, we define 
\begin{equation*}
    \bm x\cdot\bm y=\sum_{j=1}^nx^jy^j,\quad|\bm x|=(\bm x\cdot\bm x)^{1\over2}.
\end{equation*}
For any multi-index $\bm\alpha\in\mathbb Z_{\ge0}^n$, we define $|\bm\alpha|=\alpha_1+\cdots+\alpha_n$, $\bm{\partial^\alpha}=\partial_1^{\alpha_1}\cdots\partial_n^{\alpha_n}$, $\partial_j=\partial/\partial x^j$.  
The action of $R\in{\rm O}(n)$ on $\mathbb R^n$ is given by (\ref{eq:Rx}) and translation by $\bm y\in\mathbb R^n$ is given by (\ref{eq:Tyx}).  
The Euclidean motion group is generated by rotations given by ${\rm O}(n)$ and translations given by $\bm y\in\mathbb R^n$.  
With any $\lambda>0$, we associate a dilation $D_\lambda: \mathbb R^n\mapsto\mathbb R^n$ by
\begin{equation*}
    D_\lambda\bm x=\lambda\bm x\quad\hbox{for $\bm x\in\mathbb R^n$}.
\end{equation*}

The orthogonal group, space translations, and dilations are defined to act on functions $f\in{\rm C}^\infty(\mathbb R^n;\mathbb C)$ as the corresponding pull-backs:
\begin{equation*}
    (R^*f)(\bm x)=f(R\bm x),\quad(T_{\bm y}^*f)(\bm x)=f(T_{\bm y}\bm x),\quad(D_\lambda^*f)(\bm x)=f(D_\lambda\bm x).
\end{equation*}

Consider linear partial differential operators in $\mathbb R^n$ of order $m$ of the form
\begin{equation*}
    L=\sum_{|\bm\alpha|\le m}a_{\bm\alpha}\bm\partial^{\bm\alpha},
\end{equation*}
where $a_{\bm\alpha}\in{\rm C}(\mathbb R^n;\mathbb C)$.  
We say that $L$ is \underline{\it rotation invariant\/} if and only if 
\begin{equation*}
    R^*Lf=LR^*f
\end{equation*}
for any $R\in{\rm O}(n)$ and any $f\in{\rm C}^\infty(\mathbb R^n;\mathbb C)$.  
We say that $L$ is \underline{\it Euclidean motion invariant\/} if and only if $L$ is rotation invariant and \underline{\it translation invariant\/}, namely,
\begin{equation*}
    T_{\bm y}^*Lf=LT_{\bm y}^*f
\end{equation*}
for any $\bm y\in\mathbb R^n$ and any $f\in{\rm C}^\infty(\mathbb R^n;\mathbb C)$.  
We say that $L$ is \underline{\it dilation invariant\/} if and only if for any $f\in{\rm C}^\infty(\mathbb R^n;\mathbb C)\backslash\{0\}$ the following statements are equivalent:
\begin{itemize}
    \item[(i)] $Lf=0$.
    \item[(ii)] $LD_\lambda^*f=0$ for any $\lambda>0$.
    \item[(iii)] $LD_\lambda^*f=0$ for some $\lambda\not=1$.
\end{itemize}

We now summarize invariant partial differential operators under the Euclidean motion group.

\begin{theorem}[{\cite{Nomura,Shimakura}}] \label{T:2}
For any partial differential operator in $\mathbb R^n$ of order $m$ of the form
\begin{equation*}
    L=\sum_{|\bm\alpha|\le m}a_{\bm\alpha}\bm\partial^{\bm\alpha}
\end{equation*}
with $a_{\bm\alpha}\in{\rm C}(\mathbb R^n;\mathbb C)$ and $\sum\limits_{|\bm\alpha|=m}|a_{\bm\alpha}|\not=0$, the following three statements are equivalent.
\begin{itemize}
    \item[(1)] $L$ is Euclidean motion invariant.  
    Namely,
    \begin{itemize}
        \item[(i)] For any $y\in\mathbb R^n$ and any $f\in{\rm C}^\infty(\mathbb R^n;\mathbb C)$, $T_y^*Lf=LT_y^*f$.
        \item[(ii)] For any $R\in{\rm O}(n)$ and any $f\in{\rm C}^\infty(\mathbb R^n;\mathbb C)$, $R^*Lf=LR^*f$.
    \end{itemize}
    \item[(2)] All coefficients of $L$ are constants and $L$ is rotation invariant.  
    Namely,
    \begin{itemize}
    \item[(i)] For any $\bm\alpha$ with $|\bm\alpha|\le m$ and any $\bm x\in\mathbb R^n$, $a_{\bm\alpha}(\bm x)=a_{\bm\alpha}(\bm0)$.
    \item[(ii)] For any $R\in{\rm O}(n)$ and any $f\in{\rm C}^\infty(\mathbb R^n;\mathbb C)$, $R^*Lf=LR^*f$.
    \end{itemize}
    \item[(3)]  $L$ is given by a polynomial in $\triangle$ of order $[m/2]$.  
    Namely, there exists $(b_j;0\le j\le[m/2])\subset\mathbb C$ such that $b_{[m/2]}\not=0$ and $L=\sum\limits_{j=0}^{[m/2]}b_j\triangle^j$.
\end{itemize}
\end{theorem}

\begin{corollary} \label{c:3}
    Any Euclidean motion invariant partial differential operator of the second order in $\mathbb R^n$ is represented as $L=\alpha\triangle+\beta$ with $\alpha\in\mathbb C\backslash\{0\}$ and $\beta\in\mathbb C$.
\end{corollary}

\begin{corollary} \label{c:4}
    Any Euclidean motion and dilation invariant partial differential operator of the second order in $\mathbb R^n$ is represented as $L=\alpha\triangle$ with $\alpha\in\mathbb C\backslash\{0\}$.
\end{corollary}

The proof is based on the following fundamental property of rotation invariant homogeneous polynomials.

\begin{proposition}[{\cite{Nomura,Shimakura}}]\label{p:1}
Let $j\in\mathbb Z_{\ge0}$ and let $p_j$ be a rotation invariant homogeneous polynomial in $\bm x\in\mathbb R^n$ of order $j$. 
Then:
\begin{itemize}
    \item[(1)] For $j$ even, there exists a unique $b_j\in\mathbb C\backslash\{0\}$ such that $p_j(\bm x)=b_j|\bm x|^j$ for any $\bm x\in\mathbb R^n$.
    \item[(2)] For $j$ odd, $p_j(\bm x)=0$ for any $\bm x\in\mathbb R^n$.
\end{itemize}
\end{proposition}

\noindent\underline{\it Proof (\cite{Nomura})\/}

\smallskip
For $\bm x\not=0$, we define $\varphi_j$ by
\begin{equation*}
    \varphi_j(\bm x)={p_j(\bm x)\over|\bm x|^j}\quad\hbox{for $j\ge1$},\qquad\varphi_0(\bm x)=p_0(\bm x)=a_0\in\mathbb C\quad\hbox{for $j=0$}.
\end{equation*}
Then $\varphi_j$ is invariant under rotations and dilations.  
Let $\bm x\not=0$ and $j\ge1$.  
Then there exists $R\in{\rm O}(n)$ such that $\bm x=R(|\bm x|\bm e_1)$ and we have 
\begin{equation*}
    \varphi_j(\bm x)=\varphi_j\bigl(R(|\bm x|\bm e_1)\bigr)=\varphi_j(|\bm x|\bm e_1)=\varphi_j(\bm e_1).
\end{equation*}
This implies that
\begin{equation}
    p_j(\bm x)=\varphi_j(\bm e_1)|\bm x|^j
\label{eq:phijx}
\end{equation}
for any $\bm x\not=0$.  
If $j\ge1$, $p_j(\bm0)=0=\varphi_j(\bm e_1)|\bm0|^j$ and therefore (\ref{eq:phijx}) holds for any $\bm x\in\mathbb R^n$ and $j\ge1$.  
For $j=0$, we have
\begin{equation*}
    p_0(\bm x)=a_0=p_0(\bm e_1)=\varphi_0(\bm e_1)=\varphi_0(\bm e_1)|\bm x|^0
\end{equation*}
and therefore (\ref{eq:phijx}) holds for any $\bm x\in\mathbb R^n$ and $j=0$.  
We have thus proved that (\ref{eq:phijx}) hols for any $\bm x\in\mathbb R^n$ and $j\in\mathbb Z_{\ge0}$.

Let $j$ be even.  
Then the conclusion of (1) holds with $b_j=\varphi_j(\bm e_1)$.

Let $j$ be odd.  
Then the LHS of (\ref{eq:phijx}) is a polynomial in $\bm x$ and the RHS is not a polynomial in $\bm x$ unless $\varphi_j(\bm e_1)=0$.  
This shows that $p_j(\bm x)=0$ for any $\bm x\in\mathbb R^n$ if $j$ is odd. 

\begin{remark} \label{r:4}
    In [8], the proof depends on an induction argument on $j$.
\end{remark}

\medskip
\noindent\underline{\it Proof of Theorem~\ref{T:2}\/}

\medskip
For $\bm\xi\in\mathbb R^n$, we introduce $e_{\bm\xi}\in{\rm C}^\infty(\mathbb R^n;\mathbb R)$ by
\begin{equation}
    e_{\bm\xi}(\bm x)=\exp(\bm\xi\cdot\bm x)\quad\hbox{\rm for $\bm x\in\mathbb R^n$}.
\label{eq:exi-a}
\end{equation}
Then we see that $e_{\bm\xi}(0)=1$ and for any $\bm\alpha\in\mathbb Z^n_{\ge0}$,
\begin{equation}
    \bm\partial^{\bm\alpha}e_{\bm\xi}=\bm\xi^{\bm\alpha}e_{\bm\xi}.
\label{eq:partial-a}
\end{equation}
For $\bm x,\bm\xi\in\mathbb R^n$, we define
\begin{equation}
    p(\bm x,\bm\xi)=\sum_{|\bm\alpha|\le m}a_{\bm\alpha}(\bm x)\bm\xi^{\bm\alpha}.
\label{eq:pxxi-a}
\end{equation}
By (\ref{eq:exi-a}), (\ref{eq:partial-a}), and (\ref{eq:pxxi-a}), we have
\begin{equation}
    (Le_{\bm\xi})(\bm x)=p(\bm x,\bm\xi)e_{\bm\xi}(\bm x)
\label{eq:Lexi2-a}
\end{equation}
for any $\bm x,\bm\xi\in\mathbb R^n$.  
We use the equality (\ref{eq:Lexi2-a}) in the subsequent argument.

\medskip
\noindent$(1)\Rightarrow(2)$: It suffices to prove (1)(i)$\Rightarrow$(2)(i).  
The translation invariance described in (1)(i) with $u=e_{\bm\xi}$ implies
\begin{equation}
    T_{\bm y}^*Le_{\bm\xi}=LT_{\bm y}^*e_{\bm\xi}\quad\hbox{\rm for any}\quad\bm y\in\mathbb R^n.
\label{eq:TLexi-a}
\end{equation}
The LHS of (\ref{eq:TLexi-a}) at $\bm x$ is given by
\begin{equation}
    (T_{\bm y}^*Le_{\bm\xi})(\bm x)=(Le_{\bm\xi})(\bm x-\bm y)=p(\bm x-\bm y,\bm\xi)e_{\bm\xi}(\bm x-\bm y),
    \label{eq:TLexi2-a}
\end{equation}
where we have used (\ref{eq:Lexi2-a}), while the RHS of (\ref{eq:TLexi-a}) is given by
\begin{align}
    (LT_{\bm y}^*e_{\bm\xi})(\bm x)&=\sum_{|\bm\alpha|\le m}a_{\bm\alpha}(\bm x)(\bm\partial^{\bm\alpha}T_{\bm y}^*e_{\bm\xi})(\bm x)=\sum_{|\bm\alpha|\le m}a_{\bm\alpha}(\bm x)(T_{\bm y}^*\bm\partial^{\bm\alpha}e_{\bm \xi})(\bm x)\nonumber\\
    &=\sum_{|\bm\alpha|\le m}a_{\bm\alpha}(\bm x)\bm\xi^{\bm\alpha}e_{\bm\xi}(\bm x-\bm y)=p(\bm x,\bm\xi)e_{\bm\xi}(\bm x-\bm y),
\label{eq:LTexi-a}
\end{align}
where we have used (\ref{eq:partial-a}) and (\ref{eq:pxxi-a}).  
By (\ref{eq:TLexi-a}), (\ref{eq:TLexi2-a}), and (\ref{eq:LTexi-a}), with $\bm y=\bm x$, we have
\begin{equation}
    p(\bm0,\bm\xi)=p(\bm x,\bm\xi)
\label{eq:p0xi-a}
\end{equation}
since $e_{\bm\xi}(\bm0)=1$.  
Since (\ref{eq:p0xi-a}) holds for any $\bm\xi\in\mathbb R^n$, we have
\begin{equation*}
    a_{\bm\alpha}(\bm x)=a_{\bm\alpha}(\bm0)
\end{equation*}
for any $\bm x\in\mathbb R^n$ and any $\bm\alpha\in\mathbb Z_{\ge0}^n$ with $|\bm\alpha|\le m$, as required.

\medskip
\noindent$(2)\Rightarrow(3)$: We define the following polynomial $p$ in $\bm\xi$ of order $m$ by
\begin{equation}
    p(\bm\xi)=\sum_{|\bm\alpha|\le m}a_{\bm\alpha}(0)\bm\xi^{\bm\alpha}
\label{eq:pxi-a}
\end{equation}
and for $j\in\{0,1,\ldots,m\}$, we define the following homogeneous polynomial $p_j$ of order $j$ by
\begin{equation}
    p_j(\bm\xi)=\sum_{|\bm\alpha|=j}a_{\bm\alpha}(0)\bm\xi^{\bm\alpha}.
\end{equation}
Then $p$ and $L$ are represented as $p=\sum\limits_{j=0}^mp_j$ and
\begin{equation}
    L=p(\bm\partial)=\sum_{|\bm\alpha|\le m}a_{\bm\alpha}(\bm0)\bm\partial^{\bm\alpha}.
\label{eq:L-a}
\end{equation}
By (\ref{eq:partial-a}), (\ref{eq:pxi-a}), and (\ref{eq:L-a}), we obtain
\begin{equation}
    Le_{\bm\xi}=\sum_{|\bm\alpha|\le m}a_{\bm\alpha}(\bm0)\bm\partial^{\bm\alpha}e_{\bm\xi}=\sum_{|\bm\alpha|\le m}a_{\bm\alpha}(\bm0)\bm\xi^{\bm\alpha}e_{\bm\xi}=p(\bm\xi)e_{\bm\xi}.
\label{eq:Lexi-a}
\end{equation}
For any $R\in{\rm O}(n)$, we apply (\ref{eq:Lexi-a}) with $\bm\xi\mapsto{}^t\!R\bm\xi$ to obtain
\begin{equation}
    p(^t\!R\bm\xi)e_{^t\!R\bm\xi}=Le_{^t\!R\bm\xi}=LR^*e_{\bm\xi},
\label{eq:LRe-a}
\end{equation}
where we have used
\begin{equation}
    (R^*e_{\bm\xi})(\bm x)=e_{\bm\xi}(R\bm x)=\exp(\bm\xi\cdot R\bm x)=\exp(^t\!R\bm\xi\cdot\bm x)=e_{^t\!R\bm\xi}(\bm x).
\label{eq:Rexi-a}
\end{equation}
By assumption (2)(ii), we have
\begin{equation}
    R^*Le_{\bm\xi}=LR^*e_{\bm\xi},
\label{eq:RLe-a}
\end{equation}
for any $R\in{\rm O}(n)$ and $\bm\xi\in\mathbb R^n$.  
The RHS of (\ref{eq:RLe-a}) is given by (\ref{eq:LRe-a}), while the LHS of (\ref{eq:RLe-a}) is computed as 
\begin{equation}
    R^*Le_{\bm\xi}=R^*p(\bm\xi)e_{\bm\xi}=p(\bm\xi)R^*e_{\bm\xi}=p(\bm\xi)e_{^t\!R\bm\xi},
\label{eq:RLexi-a}
\end{equation}
where we have used (\ref{eq:Lexi2-a}), (\ref{eq:pxi-a}), and (\ref{eq:Rexi-a}).  
Therefore, it follows from (\ref{eq:RLe-a}), (\ref{eq:LRe-a}), and (\ref{eq:RLexi-a}), that
\begin{equation}
    p(^t\!R\bm\xi)e_{^t\!R\bm\xi}(\bm x)=p(\bm\xi)e_{^t\!R\bm\xi}(\bm x)
\label{eq:pe-a}
\end{equation}
for any $\bm x,\bm\xi\in\mathbb R^n$ and any $R\in{\rm O}(n)$.  
Evaluating (\ref{eq:pe-a}) at $\bm x=\bm0$, we have
\begin{equation}
    p(^t\!R\bm\xi)=p(\bm\xi)
\label{eq:p-a}
\end{equation}
for any $R\in{\rm O}(n)$ and $\bm\xi\in\mathbb R^n$. 
Applying (\ref{eq:p-a}) with ${\rm O}(n)\ni R\mapsto{}^t\!R\in{\rm O}(n)$, we have
\begin{equation}
    p(R\bm\xi)=p(\bm\xi)
\label{eq:p2-a}
\end{equation}
for any $R\in{\rm O}(n)$ and $\bm\xi\in\mathbb R^n$.  
By homogeneity, for any $r>0$, we have
\begin{equation}
    p(r\bm\xi)=\sum_{j=0}^mp_j(\bm\xi)r^j.
\label{eq:prxi-a}
\end{equation}
By (\ref{eq:p-a}) and (\ref{eq:p2-a}), we see that
\begin{equation}
    \sum_{j=0}^mp_j(R\bm\xi)r^j=p(rR\bm\xi)=p(R(r\bm\xi))=p(r\bm\xi)=\sum_{j=0}^mp_j(\bm\xi)r^j
\label{eq:p3-a}
\end{equation}
holds for any $r>0$, $R\in{\rm O}(n)$ and $\bm\xi\in\mathbb R^n$.  
It follows from (\ref{eq:p3-a}) that
\begin{equation}
    R^*p_j=p_j
\label{eq:Rpj-a}
\end{equation}
holds for any $R\in{\rm O}(n)$ and $j\in\{0,1,\ldots,m\}$.  
By Proposition~\ref{p:1} applied to (\ref{eq:Rpj-a}), we see that $p_j$ is given by $p_j(\bm\xi)=b_j|\bm\xi|^j$ for $j$ even and $p_j(\bm\xi)=0$ for $j$ odd.  
Therefore $p$ is represented as
\begin{equation*}
    p(\bm\xi)=\sum_{j=0}^{[m/2]}b_{2j}|\bm\xi|^{2j},
\end{equation*}
which implies
\begin{equation*}
    L=\sum_{j=0}^{[m/2]}b_j\triangle^j
\end{equation*}
if we substitute $b_{2j}\mapsto b_j$.  
We notice that $b_{[m/2]}\not=0$ since the coefficients of the highest order $m$ of $L$ do not vanish identically.  

\medskip
\noindent$(3)\Rightarrow(1)$: It suffices to prove the Euclidean motion invariance of the Laplacian $\triangle$.  
The commutativity between space derivatives and translations implies the invariance of $\triangle$ under space translations.  
For $R\in{\rm O}(n)$, we write $R=(R^j\,_k)$.
As in (\ref{eq:LL}), we have 
\begin{equation*}
    \partial_kR^*=R^j\,_kR^*\partial_j
\end{equation*}
which in turn implies
\begin{align*}
    \triangle R^*&=\delta^{jk}\partial_j\partial_kR^*=\delta^{jk}\partial_j(R^\ell\,_kR^*\partial_\ell)=\delta^{jk}R^\ell\,_k(\partial_jR^*)\partial_\ell\\
    &=\delta^{jk}R^\ell\,_k(R^i\,_jR^*\partial_i)\partial_\ell=\delta^{jk}R^\ell\,_kR^i\,_jR^*\partial_i\partial_\ell=\delta^{\ell i}R^*\partial_i\partial_\ell\\
    &=R^*\triangle,
\end{align*}
as required. 

\medskip
\noindent\underline{\it Proof of Corollary~\ref{c:3}\/}

\smallskip
Corollary~\ref{c:3} follows from directly from Theorem~\ref{T:2} with $m=2$.

\medskip
\noindent\underline{\it Proof of Corollary~\ref{c:4}\/}

\smallskip
Corollary~\ref{c:4} follows from the same argument as in the proof of Corollary~\ref{c:2}.

\begin{remark}
    Part of the proof of Theorem~\ref{T:2} depends on the choice of monomials in $\mathbb R^n$ in the original argument in \cite{Nomura,Shimakura}. 
    Here we give a proof depending the exponential functions $e_{\bm\xi}$ in a unified way.
\end{remark}

\section{Final remarks}
\label{s:final}
A similar approach to the linear partial differential operators from the point of view of symmetry is possible for the quantum mechanics \cite{HNTO}. For related subjects, we refer the reader to the references \cite{Bargmann,Barut,Hamermesh,LevyLeblond,Roncadelli,MusielakFry1,MusielakFry2}.

\section*{Acknowledgments}
We are grateful to the referee for useful comments.

H.N. is partially supported by JSPS KAKENHI Grant No.~23K03268.
T.O. is partially supported by JSPS KAKENHI Grant No.~24H00024.

\end{document}